*CORRESPONDENCE
Mohammad M. Amirian
M.amirianmatlob@dal.ca
Andrew J. Irwin
A.Irwin@dal.ca

# Extending the Monod model of microbal growth with memory

Mohammad M. Amirian[1*], Andrew J. Irwin[1*] and Zoe V. Finkel[2]

[1]Department of Mathematics and Statistics, Dalhousie University, Halifax, NS, Canada, [2]Department of Oceanography, Faculty of Science, Dalhousie University, Halifax, NS, Canada

Monod's model describes the growth of microorganisms using a hyperbolic function of extracellular resource concentration. Under fluctuating or limited resource concentrations this model performs poorly against experimental data, motivating the more complex Droop model with a time-varying internal storage pool. We extend the Monod model to incorporate memory of past conditions, adding a single parameter motivated by a fractional calculus analysis. We show how to interpret the memory element in a biological context and describe its connection to a resource storage pool. Under nitrogen starvation at non-equilibrium conditions, we validate the model with simulations and empirical data obtained from lab cultures of diatoms (*T. pseudonana* and *T. weissflogii*) and prasinophytes (*Micromonas* sp. and *O. tauri*), globally influential phytoplankton taxa. Using statistical analysis, we show that our Monod-memory model estimates the growth rate, cell density and resource concentration as well as the Droop model, while requiring one less state variable. Our simple model may improve descriptions of phytoplankton dynamics in complex earth system models at a lower computational cost than is presently achievable.

## 1 Introduction

Phytoplankton growth models are critical components of ocean ecosystem models that link the supply of inorganic resources and light to the primary production of particulate organic material, enabling predictions of biomass distributions, resource contributions to food webs, and biogeochemical fluxes. Growth models are generally formulated in terms of resource acquisition, biomass accumulation, and cell division, but the complexity of these phenomena must be greatly simplified to be incorporated in an ecosystem model. Models require choices about the appropriate number of state variables and parameters and each investigator must select functions to describe key processes. Specific models are designed for a particular purpose, including their ability to represent key phenomena, and simpler models are usually preferred over more complex ones if they capture the phenomena of interest.

The Monod model, which is widely used in biogeochemical models, describes phytoplankton growth as a function of extracellular resource concentrations and assumes fixed cellular composition (Fasham et al., 1990; Follows and Dutkiewicz, 2011; Aumont et al., 2015; Dutkiewicz et al., 2020; Henson et al., 2021; Follett et al., 2022). In this approach, growth rate is a function of resource concentration defining whether resources are limiting or in excess, according to the half-saturation constant, and a maximum growth rate, or equivalently a maximum resource uptake rate. Resources are converted into biomass at a constant yield. A consequence of the constant yield is that the model does not describe an internal physiological state of cells. The model may perform poorly when the resource concentration changes rapidly relative to the desired time scale of prediction as it does not account for resources taken into the cell which are not yet used for growth or the possibility of variable cell quotas (Moore et al., 2001; Flynn, 2010; Akoglu, 2020; Tsakalakis et al., 2022). When predictions are computed over time scales that are long relative to external fluctuations and cell generation times, this simplification may have little effect, but internal storage, acclimation, and external fluctuations are known to be important for determining population size, competitive ability, and stoichiometry (Tozzi et al., 2004; Finkel et al., 2007; Finkel et al., 2010; Wu et al., 2014; Irwin et al., 2015).

Phytoplankton vary their composition in response to changing resource conditions, which can be described by the more complex Droop model (Droop, 1968). The Droop model is a widely-used solution to many problems that may arise using the constant quota Monod model. Droop model adds a second state variable per species to track the internal physiological state of the cell, described as a variable cell quota (Pahlow and Oschlies, 2013). The effect of variable internal storage is to record a memory of past environments to the cell, so that growth rate does not simply depend on the current external nutrient conditions. In the Droop model, resources are taken into the cell as in the Monod model, but the growth rate depends on resources inside the cell rather than the extracellular resources in the Monod model. This added complexity allows phytoplankton described by the Droop model to better respond to changing conditions. The cost is a doubling of the number of state variables per species and the need to provide additional parameters to describe the dynamics of this internal resource pool. Although the parameters in the Monod and Droop models are straightforward to interpret, they are not necessarily easy to quantify. The key parameters minimum cell quota ($Q_{min}$) and maximum uptake rate ($V_{max}$) are extrapolations from a series of experiments and can't be directly observed, which may cause mis-estimation of parameter values (Shuter, 1978; Irwin et al., 2006; Mei et al., 2009; Finkel et al., 2016). Even the maximum growth rate presents challenges, as it is difficult to know in a laboratory if all the relevant experimental conditions are optimal for growth. Common solutions to these challenges are to estimate parameters from a combination of lab data and numerical tuning exercises to suit an application (Irwin and Finkel, 2018) or identify successful species from ecological simulations (Follows et al., 2007). Some researchers have explored the use of variable quotas in biogeochemical models by a parameterization of uptake (Moore et al., 2001) and agent-based models of individual cells instead of population-level averages (Hellweger and Kianirad, 2007). Incorporating variable quota has an impact on stoichiometry, productivity, and ecosystem-level biogeochemical cycles, but the consensus appears to be that the effect is minor relative to the computational cost and model complexity (Kwiatkowski et al., 2018; Anugerahanti et al., 2021). The Monod and Droop equations are two fundamental models that describe the phytoplankton growth function, but other models have been developed to study the effects of multiple limiting resources. One approach subdivides the cell into compartments each with distinct resource costs and roles in resource acquisition and growth (Shuter, 1979). These models have been used to describe the effect of co-limitation of light and nitrogen on growth or nitrogen and iron on diazotrophy as well as the effects of temperature on the metabolic activity and cellular composition (Geider et al., 1998; Nicholson et al., 2018). These models are considerably more complex than the simpler approach taken by Monod, Droop and our new model presented here.

We propose a new model for phytoplankton growth, which combines characteristics of the Monod and Droop models. Our model combines the fixed composition of the Monod model with a more complex growth function that approximates a memory of the past history of extracellular conditions, allowing the growth rate to change as resources are depleted and mimicking the predictions of the Droop model. We use a fractional calculus extension of ordinary differential equations to include the past, in acclimation time scale, in addition to the current state of the system into the dynamical equations (Matlob and Jamali, 2019; Eftekhari and Amirian, 2022). Here we provide a theoretical justification for the model formulation, explore the mathematical, physical and biological interpretation of its parameters and equilibrium conditions, suggest a way to measure the cell memory empirically, and demonstrate its performance compared to Monod and Droop models with both simulations and experimental data from the laboratory.

## 2 Methods

The Monod model describes the growth rate of microorganisms, $\mu$ ($d^{-1}$), as a saturating function of the concentration of the limiting resource in the growth medium, $R$ (mol $L^{-1}$),

$$\mu(R) = \mu_{max}\left(\frac{R}{K_\mu + R}\right) \qquad (1)$$

where $\mu_{max}$ ($d^{-1}$) is the asymptotic maximum growth rate and $K_\mu$(mol $L^{-1}$) describes the concentration at which the growth rate is half its maximum. While the growth of cells is a complex process arising from a multitude of enzymatic processes, a common approach is to describe growth as an enzymatic process using the Michaelis-Menten model with a single resource as substrate. This parameterization approximately describes growth rates at a range of steady-state nutrient concentrations (Henriques and Balsa-Canto, 2021). Many modifications and extensions have been added, including changes to describe growth rate inhibition at high resource concentrations and the costs of maintenance metabolism (Kovarova-Kovar and Egli, 1998; Lee et al., 2015). Notation for all models is summarized in Table 1.

The Droop model describes growth as a two-step process where resources are first taken into the cell and then used for growth (Droop, 1968; Grover, 1991; Verdy et al., 2009). Resource uptake, $\rho$ (mol cell$^{-1}$ d$^{-1}$), is commonly written with the same functional form as growth rate in the Monod model,

$$\rho(R) = V_{max}\left(\frac{R}{K_\rho + R}\right) \qquad (2)$$

where $V_{max}$ (mol cell$^{-1}$ d$^{-1}$) is the asymptotic maximum uptake rate and $K_\rho$ (mol $L^{-1}$) describes the concentration at which the uptake rate is half its maximum. Resources taken into the cell form an internal pool, called the quota, $Q$ (mol cell$^{-1}$), which is used to predict the growth rate. The growth rate is a saturating function shifted to the right to establish a minimum cell quota, $Q_{min}$ (mol cell$^{-1}$), at which growth rate is 0,

$$\mu(Q) = \mu_{max}\left(1 - \frac{Q_{min}}{Q}\right) \qquad (3)$$

This two-step process partially decouples growth from resource uptake. In particular, unlike the Monod model, this formulation allows for continued growth for a time after external resources are depleted. The internal resource pool acts as the cell's memory in the sense that the growth rate now depends on stored resources and not on the concentration of resources outside the cell. The Droop model of growth allows for a variable quota in contrast to the constant quota (or yield of cells produced per resource consumed) of the Monod model. The minimum cell quota can be estimated from a regression of cell quota over a wide range of growth rates under balanced growth conditions. The minimum quota is determined by extrapolating the regression line to zero growth rate, so it is desirable to have quota estimated from some cultures with very slow growth rates, which can take considerable experimental effort. Balanced growth at slow growth rates and low cell quotas can be difficult to maintain, making minimum cell quota challenging to quantify experimentally, while the average yield is much easier to estimate.

Both the Monod and Droop models are typically used to describe balanced growth, such as observed in a continuous culture at equilibrium. The Droop model describes some non-equilibrium growth rates well, including resource starvation and resupply, even under conditions where the Monod model performs poorly. Here we develop a simple alternative model that describes growth rate under changing resource conditions as well as balanced growth. Our model formulation is motivated by the storage effect of Droop's variable quota. We use a memory analogy and make growth rate dependent on past resource conditions. We do this in a particularly simple way to arrive at a parameterization of the growth rate with no explicit dependence on past conditions. Our formulation modifies and extends the concept of half-saturation constant to incorporate the impact of changing uptake rate and quota needed for cell division that occurs over nutrient concentrations ranging roughly from $K_\mu$ to $K_\rho$, sometimes called the acclimation range (Morel, 1987; Smith et al., 2009).

We reinterpret growth rate in the Monod model (Eq. 1) as the solution of a differential equation describing the change in growth rate with the change in the dimensionless resource-availability ratio $N = R/(K_\mu + R)$,

TABLE 1 List of state variables and parameters in Monod, Monod-memory, and Droop models.

| | Symbol | Units | Description |
|---|---|---|---|
| State variables | $R$ | mol $L^{-1}$ | Resource concentration |
| | $Q$ | mol cell$^{-1}$ | Cell quota |
| | $X$ | cell $L^{-1}$ | Cell density |
| Parameters | $\mu_{max}$ | d$^{-1}$ | Maximum growth rate |
| | $K$ | mol $L^{-1}$ | Saturation constant for growth ($K\mu$, Monod, $K$, Monod-memory) or uptake ($K_\rho$, Droop) |
| | $\alpha$ | – | Memory element for growth function (Monod-memory) |
| | $V_{max}$ | mol cell$^{-1}$ d$^{-1}$ | Maximum uptake rate (Droop) |
| | $Q_{min}$ | mol cell$^{-1}$ | Minimum cell quota (Droop) |

$$\frac{d\mu}{dN} = \mu_{\max} \qquad (4)$$

The ratio $N$ is always between 0 and 1. The initial condition $\mu$ ($N = 0$) $= 0$ ensures that growth rate is 0 when no resources are available. The solution is $\mu = \mu_{\max}N$ in agreement with Eq. (1). Growth rate increases asymptotically to its maximum as $R \rightarrow \infty$ ($N \rightarrow 1$, $\mu(1) = \mu_{\max}$). A non-zero growth rate at $N = 0$ can be accommodated by changing the initial condition resulting in models given in (Marr et al., 1963; Pirt, 1965; Van Uden, 1967; Boethling and Alexander, 1979; Rittmann and McCarty, 1980; Sancho et al., 1997).

We add a memory of past conditions by modifying this differential equation. We write it in the form of an integral equation, $\mu(N) = D_{0,N}^{-1}\mu_{\max} + \mu(0)$, then generalize the integration to the Riemann-Liouville fractional integral, defined for a function $f$ as follows,

$$D_{0,t}^{\alpha}f(t) = \frac{1}{\Gamma(\alpha)} \int_{0}^{t} (t-\tau)^{\alpha-1} f(\tau) d\tau \qquad (5)$$

where $\alpha > 0$ and $\Gamma$ is the gamma function (Matlob and Jamali, 2019). Note that this definition is equal to the ordinary Riemann integral when $\alpha = 1$. After making this change Eq. (4) is transformed to

$$\mu(N) = D_{0,N}^{-\alpha}\mu_{\max} + \mu(0), \quad \alpha > 0 \qquad (6)$$

Setting $0 < \alpha \le 1$, the Riemann-Liouville operator defines a convolution of $\mu_{\max}$ with a power-law weight function $N^{\alpha-1}/\Gamma(\alpha)$. This function gives more weight to $\mu_{\max}$ as the resource-availability ratio goes toward zero, $N \rightarrow 0$, such that the weighted sum increases with decreasing $\alpha$ (Figure S1), modulating growth rate in the face of nutrient limitation and starvation. The growth rate depends on an average of resource concentrations outside the cell, including potentially less-limiting concentrations from the recent past, instead of just current resource concentrations outside the cell. We refer to $\alpha$ as the memory element (or cell memory) following previous interpretations in the literature (Amirian et al., 2020; Khalighi et al., 2022). The appropriate value of $\alpha$ for a particular species will be determined by statistical data analysis incorporating a population model, presented below. We restrict the value of $\alpha$ to the interval (0, 1]. If $\alpha$ is greater than 1, the qualitative change in the weight function in Eq. (5) alters its interpretation so that the memory concept does not hold.

The integral equation (Eq. 6) can be solved explicitly (Matlob and Jamali, 2019), so the memory of past nutrient concentrations can be included in the growth function without requiring the computational burden of tracking past nutrient conditions. The solution and our proposed Monod-memory model for growth rate is

$$\mu(R) = \frac{\mu_{\max}}{\Gamma(1+\alpha)} \left( \frac{R}{K+R} \right)^{\alpha}, \quad 0 < \alpha \le 1 \qquad (7)$$

where we have assumed a growth rate of zero when resources are completely depleted, $\mu(0) = 0$. This growth function is very similar to the original Monod function, with two changes. A scaling factor of $1/\Gamma(1 + \alpha)$ inflates the realized maximum growth rate relative to the value of $\mu_{\max}$ by up to roughly 13% when $\alpha \approx 0.45$ and has almost no influence on $\mu_{\max}$ when $\alpha$ is approximately 0 or 1 (Figure S2). The memory element $\alpha$ reshapes the original saturating function of the resource availability ratio (Figure S3). As a result, the constant $K$ is no longer a half-saturation constant: where $R = K$, resource limitation reduces the growth rate by a factor $2^{-\alpha}$, which is larger than 1/2 when $0 < \alpha < 1$. Half-saturation of growth rate ($\mu = \mu_{\max}/2$) occurs when the resource concentration is given by

$$R = \left[ \left( \frac{2}{\Gamma(1+\alpha)} \right)^{1/\alpha} - 1 \right]^{-1} K \stackrel{\text{def}}{=} M(\alpha)K \qquad (8)$$

meaning that $K$ decreases with increasing $\alpha$ due to the increase in the memory function, $M(\alpha)$, decreasing the uptake rate accordingly. We name the $K$ parameter the memory-half-saturation constant, as the acclimation range concept is incorporated in the growth function. If $\alpha = 1$, half the maximum growth rate is achieved at $R = K$, which means $K$ is then the half-saturation constant described in the Monod model.

## 2.1 Population dynamics

To describe the change in a population over time, we embed the growth rate equations into a set of differential equations describing the dynamics of resource concentration and cell density. The state variables are resource concentration in the growth medium $R$ (mol $L^{-1}$), cell quota $Q$ (mol cell$^{-1}$), and cell density $X$ (cells $L^{-1}$). For both the Monod and Monod-memory growth functions we assume a constant cell quota. Fresh media is supplied at a rate $d \ge 0$ (d$^{-1}$) with resource concentration $R_s$(mol $L^{-1}$). Media, including cells and resources at well-mixed concentrations are washed out at the same rate. The system of equations is

$$\frac{dR}{dt} = d(R_s - R) - \rho X$$
$$\frac{dX}{dt} = (\mu - d)X. \qquad (9)$$

The Monod and Monod-memory models do not model quota or resource uptake explicitly. The quantity of resource consumed by the production of new cells ($\rho X$) is typically described in terms of the yield or its reciprocal the quota. We describe resource depletion by growing cell populations as resource uptake as described in the Droop model and note this is equal to resource consumption (growth rate times quota) at equilibrium ($\rho = \mu Q$), where $V_{\max} = \mu_{\max}Q$. The Monod-memory model has a different expression for the growth rate that incorporates the memory element, $\alpha$, regulating the growth against nutrient limitation or starvation so that the cell has the opportunity to grow for a time after the resources have been depleted.

We contrast the fixed-quota Monod and Monod-memory models with the variable internal stores Droop model (Grover, 1991), which adds an equation to describe the change in cell quota resulting from the combination of the uptake of resources from the media $\rho$ (mol cell$^{-1}$ d$^{-1}$) and consumption from growth,

$$\frac{dR}{dt} = d(R_S - R) - \rho X$$
$$\frac{dQ}{dt} = \rho - \mu Q \qquad (10)$$
$$\frac{dX}{dt} = (\mu - d)X.$$

Here the growth rate depends on cell quota (Eq. 3) and resource uptake into the internal storage pool is a Michaelis-Menten function (Eq. 2).

There are three different equilibrium points, depending on the dilution rate. In a batch culture with no dilution ($d = 0$) the equilibrium state is $R = 0$, $\mu = 0$ and the final number of cells is a function of the initial conditions. With a large dilution rate ($d > \mu_{max}$), washout occurs and no cells remain ($X^* = 0$, $R^* = R_s$). Intermediate dilution rates lead to equilibrium rates ($R^*, X^*, Q^*$) depending on the parameters and the model. For Droop model

$$R^* = \left(\frac{\mu_{max}(Q^* - Q_{min})}{V_{max} - \mu_{max}(Q^* - Q_{min})}\right)K,$$

$$X^* = \frac{d}{\rho}(R_s - R^*), \qquad Q^* = \frac{\mu_{max}}{\mu_{max} - d}Q_{min}$$

For Monod-memory model

$$R^* = \left(\frac{(d\Gamma(1 + \alpha)/\mu_{max})^{1/\alpha}}{1 - (d\Gamma(1 + \alpha)/\mu_{max})^{1/\alpha}}\right)K, \qquad X^* = \frac{d}{\rho}(R_s - R^*)$$

The equilibrium for the Monod model is the same as the Monod-memory model with $\alpha = 1$.

## 2.2 Simulated populations

*Batch Culture*. We simulated a batch culture (no dilution with fresh media, $d = 0$) using the Droop model (Eq. 10) with initial values for resource concentration, cell quota, cell density ($R_0, Q_0, X_0$) = (2.5,2,2.5) and parameters ($\mu_{max}, V_{max}, Q_{min}, K, d$) = (2,3,0.1,10,0). We multiplied the output ($R, Q, X, \mu$) by the exponential of independent samples from a Normal random variable with mean 0 and variance 0.1 to simulate measurement error and avoid a perfect match to the Droop model. These quantities are all positive, so the exponential of a Normal random variable was used to avoid generating non-physical negative values. Simulated data plus measurement error are shown as red dots in Figure S4. Parameters for all three models were estimated by maximum likelihood estimation using these simulated data.

*Double-Batch Culture*. To test model performance when resources were resupplied, we extended the batch culture simulation using the same parameters and initial state variables. When cell quota was close to its asymptotic stable value, $Q_{min} = 0.1$, we re-supplied the same amount of nutrient from the initial culture, $R_0 = 2.5$, in a pulse maintaining zero dilution rate. We included the same amount of error, $N(0, 0.1)$, to the generated data as used for the batch culture simulation.

## 2.3 Experimental data

We tested the three population models with data obtained from four phytoplankton cultures grown in the lab: diatoms (*Thalassiosira pseudonana* and *T. weissflogii*) and prasinophytes (*Micromonas* sp. and *Ostreococcus tauri*). These species vary in cell volume by roughly a factor of 1000 from the smallest photosynthetic eukaryote (*O. tauri*, 1.8 $\mu$m$^3$) to a medium-sized diatom (*T. weissflogii*, 1630 $\mu$m$^3$). Larger cells, particularly diatoms, are known to have greater capacity to store 'surplus' resources, in terms of their total mass and relative to their minimum cell mass, compared to smaller cells. Cultures were acclimated to replete conditions for 8 generations and then transferred to N-free media to induce nitrogen stress and followed in a batch culture (Liefer et al., 2019). Observations of cell density (cells L$^{-1}$), dissolved inorganic nitrogen in the media ($\mu$mol L$^{-1}$), and cell N quota (mol cell$^{-1}$) were collected during nutrient-replete, balanced exponential growth ($t = 0$) and four additional points spanning the late-exponential and stationary phase. Growth rate (d$^{-1}$) was computed from daily cell counts in steady-state exponential growth and for each sampling day in the N-starved batch culture. Further experimental details and results are available in (Liefer et al., 2019). We analyzed the data from each replicate and each species separately and computed average parameters from the three replicates.

## 2.4 Quantifying the memory element

Here we give a biological meaning to the memory element, $\alpha$, and suggest a way to calculate it from experimental data through equating our Monod-memory growth function (Tsakalakis et al., 2022) to Droop's (Aumont et al., 2015) and solving for the $\alpha$ parameter. We scale the maximum growth rate in Droop, $\mu_{max}$, to be equal to the fraction $\mu_{max}/\Gamma(1 + \alpha)$ in our model, obtaining

$$\left(\frac{R}{K + R}\right)^\alpha \approx \left(1 - \frac{Q_{min}}{Q}\right) \qquad (11)$$

We estimate an average value for $\alpha$ by noting that $Q$ and $R$ are state variables generated by population dynamics models, taking logs and finding the expectation over the computed time series, obtaining

$$\alpha \approx \frac{E\left[\log\left(1 - \frac{Q_{min}}{Q}\right)\right]}{E\left[\log\left(\frac{R}{K+R}\right)\right]} \tag{12}$$

where $E$ stands for the expectation. Using the Taylor expansion $\log(1-x) = -\sum_{n=0}^{\infty} \frac{x^n}{n}$, with an $O((\frac{Q_{min}}{Q})^2)$ error, $\alpha \propto E[Q]^{-1}$. Rewriting the fraction part of Eq. (12) in the form of uptake function (Follett et al., 2022), we can see $\alpha \propto E[\log(\frac{\rho(R)}{V_{max}})]^{-1}$. The memory element, $\alpha$, depends on both cell quota, $Q$ and uptake rate $\rho(R)$, coupling the internal storage information to the current external nutrient in the environment. The values of $Q$ and $\rho$ are positively correlated (since at equilibrium $\rho = \mu Q$) so $\alpha$ decreases with increasing $Q$ (Verdy et al., 2009). In general, larger cells have greater uptake rates and longer memory (Eq. 12). For cells with short memory ($\alpha$ close to 1), the reverse is true. Our formulation of the growth function enables us to transfer the impact of internal storage into Monod's model with the memory element, $\alpha$. This preserves some of the information about cell storage described by variation in $Q$ when we project from a 3-dimensional space $(X,Q,R)$ to a 2-dimensional plane $(X,R)$. This agrees with the commonly known interpretation of the Riemann-Liouville fractional integral in mathematics, projecting function information from a higher to lower dimension (Podlubny, 2002) and explains our reasoning for selecting this operator.

The memory element, $\alpha$, can be calculated by Eq. (12), the ratio of the mean log of growth rate over maximum growth rate to the mean log of uptake rate over maximum uptake rate. To emphasize this interpretation we rewrite the equation as

$$\alpha = \frac{\text{mean } \{\log(Growth\ Rate\ over\ Maximum\ Growth\ Rate)\}}{\text{mean } \{\log(Uptake\ Rate\ over\ Maximum\ Uptake\ Rate)\}} \tag{13}$$

We seek a simple way to modify the Monod model so that our new model explains the empirical data as well as the Droop model. Our approach exploits a simple but broad relationship between the memory element and physiological changes in the cell. One may calculate the value of $\alpha$ from either the steady-state (Burmaster, 1979) or short-term case (Morel, 1987; Smith et al., 2009). In such scenarios we may expect to derive a somewhat different value than our general formula (Irwin et al., 2015), as the acclimation range could be different in these cases (Morel, 1987; Smith et al., 2016; Ward et al., 2017).

To test the above equation, we performed 200 independent batch-cultured simulations using the variable quota Droop model with parameters and initial values drawn from ranges given in Table S1 for each taxon. For each replicate, we calculated the value of $\alpha$ using Eq. (13). Then we applied our statistical method to each replicate individually to estimate the parameters for our Monod-memory model (Eq. 7) using maximum likelihood. We scaled the state variables of the Monod-memory model, $R$ and $X$, to unit variance to give equal weight to each in the maximum likelihood calculation. For a few simulations (less than 10%) the optimization did not converge, so they were not considered further.

## 2.5 Statistical analysis

We estimated parameters for models using maximum likelihood estimation with a loss function in the residual function form, $L(\theta, \hat{\theta}) = |\theta - \hat{\theta}|$, following the standard normal distribution with zero mean, where $L$, $\theta$ and $\hat{\theta}$ are log-likelihood, the observational data and predicted values, respectively. The optim function (Nocedal and Wright, 2006) was used to maximize the log-likelihood function using R statistical software (R version 4.0.5). We used deSolve function with ode45 method (Soetaert et al., 2010) to solve the ordinary differential equations numerically as part of this computation. We compared the performance of all models using mean squared error (MSE).

# 3 Results

## 3.1 Batch-culture simulations

We tested the population dynamic models with constant (Eq. 9) and variable (Eq. 10) quota, estimating model parameters (Table S2) and mean squared error (MSE) for each state variable (Table S3). Monod's model (blue line) fails to explain changes in the growth rate and substrate concentration, while the Droop variable quota model (green line) and Monod-memory model (black line) closely follow the trend in the data (red dots, $n = 30$, Figure 1 for $t< 5$, Figure S4). The MSE for the Droop model, which was used to generate the simulated data, is somewhat smaller than the MSE for the Monod-memory model; both are clearly superior to the MSE for the Monod model.

We extended the Droop model batch simulation by resupplying resources in a pulse in stationary phase at time $t = 5$ to increase $R$ to the same concentration as the initial conditions (red dots, $n = 50$, Figure 1). The Monod-memory model tracked the resource substrate concentration as well as the Droop model. The Monod-memory predicted cell density and growth rate that were too large relative to the simulation and Droop model, with an error comparable to the Monod.

The simulated and predicted increase in cell quota (Droop model) were in agreement, but quite small relative to the initial phase of the batch culture ($t< 2$).

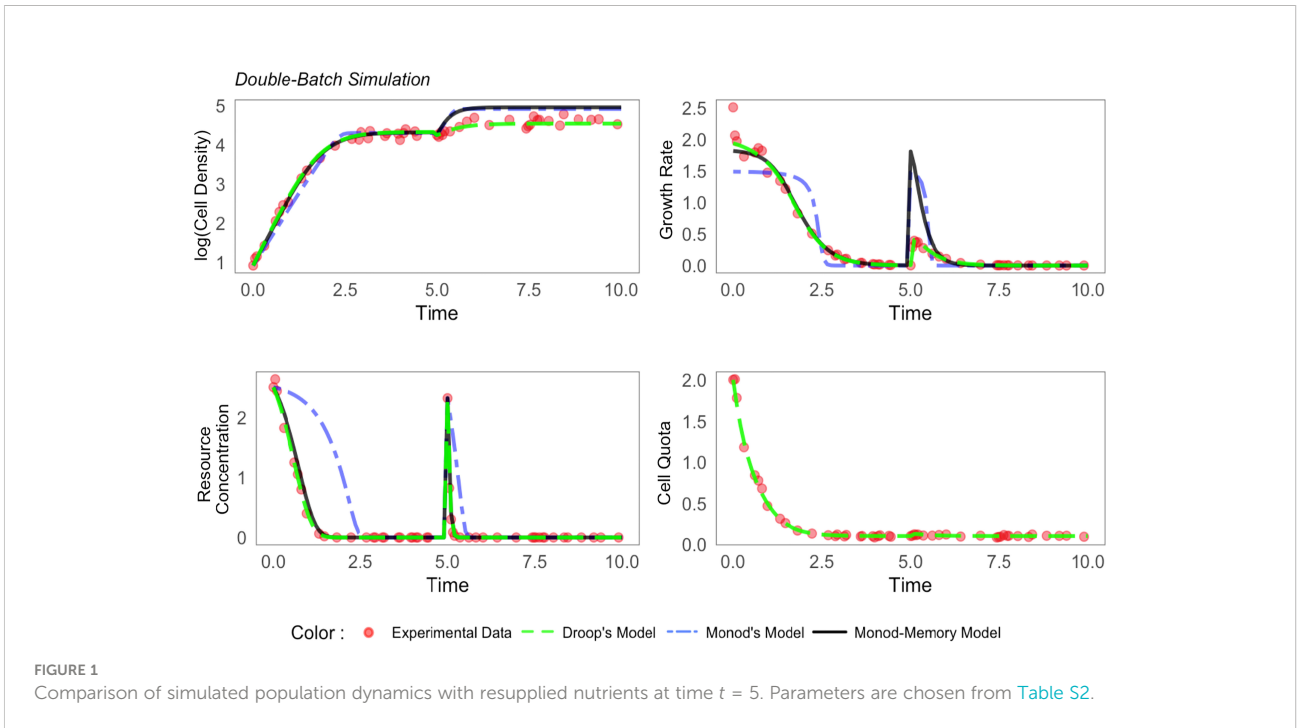

**FIGURE 1**
Comparison of simulated population dynamics with resupplied nutrients at time $t$ = 5. Parameters are chosen from Table S2.

## 3.2 Analysis of empirical data

Echoing results from the simulations, Monod's model failed to explain important variation in the growth rate and substrate concentration for all four species (Figures 2 and S5-S7). For

*Thalassiosira pseudonana* (Figure 2) and *Micromonas* sp. (Figure S6), Monod-memory and Droop models performed similarly in estimating the substrate concentration, cell density, and growth rate. Despite estimating substrate concentration and cell density slightly better (Table S5), the Droop model overestimated the maximum growth rate, while this is not the case in our Monod-memory model (Table S4). Our model predicted growth rate better than the Droop model. For *Ostreococcus tauri* (Figure S7), the Monod-memory model resulted in better estimates of substrate concentration, cell density, and growth rate compared to Droop's model (Table S5), while the reverse is the case for *Thalassiosira weissflogii* (Figure S5). In both cases, the Droop model overestimated the maximum growth rate (Table S4).

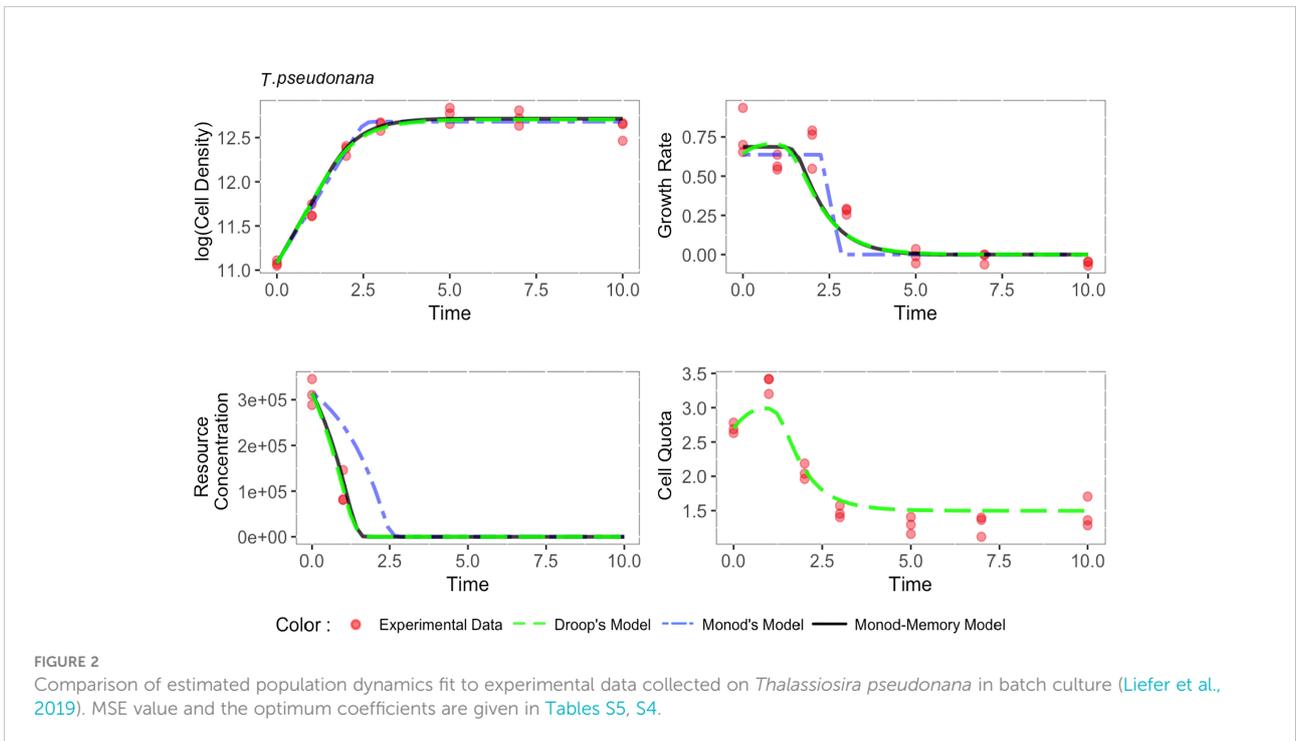

**FIGURE 2**
Comparison of estimated population dynamics fit to experimental data collected on *Thalassiosira pseudonana* in batch culture (Liefer et al., 2019). MSE value and the optimum coefficients are given in Tables S5, S4.

## 3.3 Quantifying the memory element

The $\alpha$ parameter can be estimated either empirically or statistically (Eqs. 12, 13). Both methods yielded similar results with only minor deviations from each other (Figure S8). Estimated mean values and standard deviation are given in Table 2. Cell memory depends upon cell size (Figure 3 and Table S4). Cells with small values of $\alpha$ have long memories and continue to grow at large fractions of their maximum growth rate long after external nutrients have been drawn down to lower levels. Thus, taxonomic differences between prasinophytes and diatoms may be reflected in estimated parameter values, particularly the memory element $\alpha$ as a consequence of differences in cell volumes. The correlation matrix for diatoms (*T. pseudonana* and *T. weissflogii*) and prasinophytes (*Micromonas* sp. and *O. tauri*), showed that the model parameters correlation matrix for diatoms was different compared to prasinophytes (Figure 4). Cell memory was negatively correlated with maximum uptake rate in smaller-sized prasinophytes, while this correlation was positive (or zero) in the relatively larger-sized diatoms.

The maximum growth rate was estimated to be two times more than the value measured in the lab for *T. pseudonana* and *T. weissflogii* and three times more for *Micromonas* sp., and *O. tauri* using the Droop model. Unlike Droop's results, these values are close to each other for our model (Figure S9), so in this context, it seems that the Monod-memory model provided a better estimate of $\mu$max.

## 4 Discussion

Our model combines the relative simplicity of the Monod model with the added flexibility of the Droop model. We obtained our formulation by modifying a differential equation to use a fractional integration, so that resource uptake and growth depends on the change of resource concentration outside the cell. This effectively creates a memory of past nutrient concentrations without requiring the computational burden of tracking past nutrient conditions. Growth rates change as resources are depleted, mimicking the predictions of the Droop model and describing the transition from limiting to saturating nutrient conditions better than the Monod model. Growth doesn't stop as soon as resource concentration becomes strongly limiting because of an implicit storage pool in the model. This pool is not modeled over time, but its average impact on growth rate is estimated by the memory parameter, $\alpha$.

Our model simplifies the evolution of the system from a trajectory in the 3-dimensional space, $(X,Q,R)$ by projection into the 2-dimensional plane $(X,R)$. The effect of the projection on the dynamical system is described by the memory parameter $\alpha$ and the Riemann-Liouville fractional integral operator. This is a practical application of the geometric interpretation for the Riemann-Liouville operator proposed by Igor Podlubny which projects the dynamical behavior of a system from a higher to a lower dimensional space (Podlubny, 2002). For $0 < \alpha \leq 1$, the power-law weight function in the Riemann-Liouville convolution integral gives more weight to the maximum growth rate, $\mu_{max}$ as the

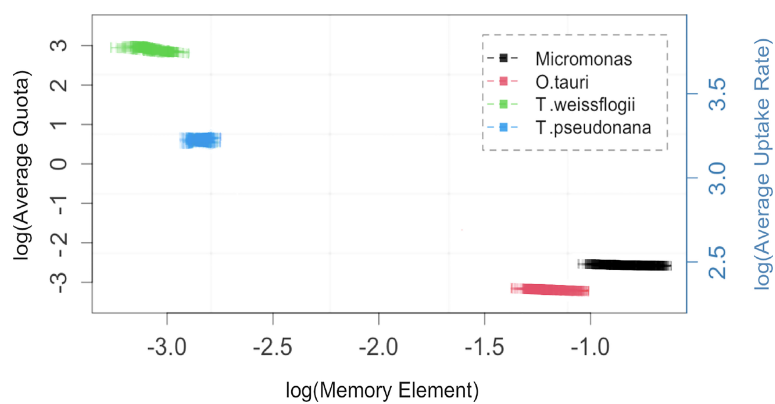

FIGURE 3
The memory element, $\alpha$, (squares with 95% confidence interval) is negatively correlated with average quota and uptake rate for the four phytoplankton taxa studied. Replicates for each tax a are generated from 200 independent fits of the Monod-memory model to Droop model with parameters and initial values drawn from ranges given in Table S1 for each taxon. Note natural log scales on both axes.

TABLE 2  Estimated mean values and standard deviation for the Monod-memory (Eq. 7) parameters.

| Taxon | $\hat{\alpha} \pm$ SD | $\hat{\mu}_{max} \pm$ SD | $\hat{K} \pm$ SD | $\hat{V}_{max} \pm$ SD |
|---|---|---|---|---|
| T. pseudonana | 0.058 ± 0.0012 | 0.704 ± 0.0251 | 0.125 ± 0.0063 | 3.005 ± 0.1074 |
| T. weissflogii | 0.046 ± 0.0027 | 0.567 ± 0.0163 | 0.102 ± 0.0099 | 2.236 ± 0.0654 |
| O. tauri | 0.303 ± 0.0250 | 0.753 ± 0.0198 | 0.344 ± 0.0371 | 2.038 ± 0.0142 |
| Micromonas | 0.440 ± 0.0448 | 0.571 ± 0.0159 | 1.266 ± 0.1308 | 2.355 ± 0.1537 |

A ˆ indicates the value was estimated by maximum likelihood.

resource-availability ratio decreases such that the weighted sum increases with decreasing $\alpha$, modulating the growth in the face of nutrient limitation or starvation (Figure S1). We quantified the memory parameter by statistical parameter estimation using the dynamical model for a continuous culture. We showed that this estimate can be approximated by the simpler and intuitive ratio of the mean log of growth rate divided by maximum growth rate to the mean log of uptake rate divided by the maximum uptake rate, computed over the duration of the experiment (Eq. 13, Figure S8).

The Monod model fails to explain transient changes in growth rate under nutrient starvation and limitation as it equates resource uptake and resource used for growth (Anugerahanti et al., 2021). The Droop and Monod-memory models both estimated the cell density, substrate concentration and the growth rate well. However, our Monod-memory model with constant quota estimated the growth rate 8% and 21% better for *Micromonas* sp. and *T. pseudonana*, respectively (Table S5). In all cases, the Droop model overestimated the maximum growth rate, by a factor of approximately 2 compared to than the maximum value measured in the lab for *T. pseudonana*, *T. weissflogii* and approximately 3 for *Micromonas* sp. and *O. tauri*. We did not observe this overestimation with the Monod-memory model. This overestimation is a known consequence of the Droop formulation, where the maximum growth rate is achieved asymptotically as $Q \to \infty$. For this reason, some researchers call the parameter $\mu'_{max}$) (Grover, 1991). In an environment where the resource concentration changes rapidly relative to the desired time scale of prediction, the Monod model approximation is only reasonable if the cell has a very small storage to rely on for growth. Despite these limitations, the Monod model is usually preferred over the more complex Droop model, as the implementation of the Droop model presents difficulties in large ecosystem simulations due to a doubling of the computational cost and difficulty in defining the equations to track changes in inter-cellular nutrient concentrations (known as the challenge of advecting quotas) (Kwiatkowski et al., 2018).

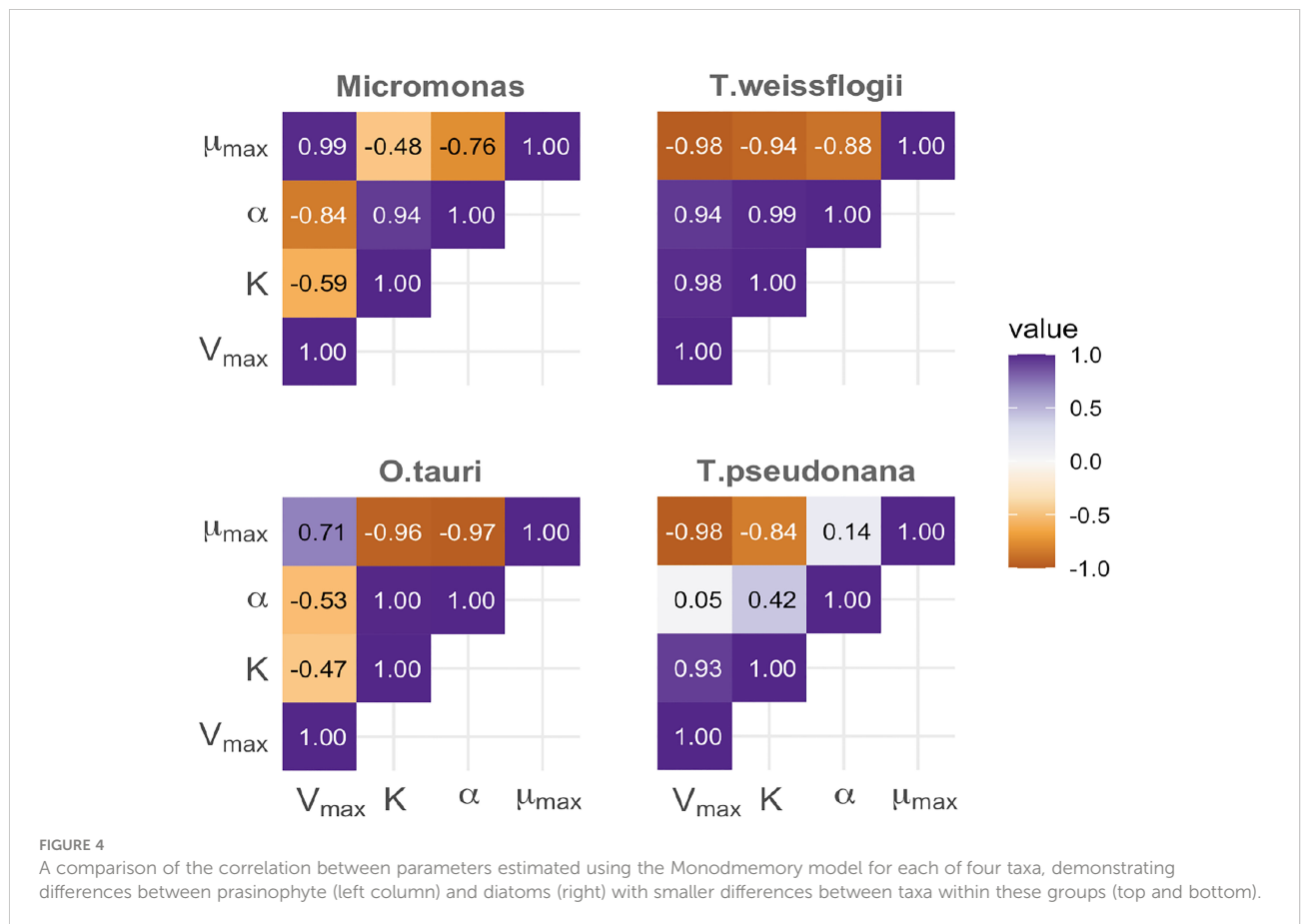

FIGURE 4
A comparison of the correlation between parameters estimated using the Monodmemory model for each of four taxa, demonstrating differences between prasinophyte (left column) and diatoms (right) with smaller differences between taxa within these groups (top and bottom).

Our Monod-memory model does not track changes in cell quota, $Q$, instead it relies on the average $Q$ to convert resources into cells. If total mass is computed from resources in the media, cell density, and average $Q$, the model will not conserve mass. Estimated cell quota can be computed for any time in the simulation and used to compute total mass of the system. The Monod-memory growth function (Eq. 7) is a consistent estimator of the Droop growth function (Eq. 3), yielding a reasonable reconstruction of variable quota over time by solving Eq. (12) for $Q$. Mass is conserved if we use the computed variable quota instead of an average quota.

The Monod-memory growth equation is a modified hyperbolic Monod-type function of extracellular resource concentration, combining the characteristics of the Monod and Droop models, thereby taking the cell size concept into account without the extra computational cost of the Droop model. Our mathematical, physical and biological interpretations show that cells with larger storage pools have longer memory (smaller $\alpha$). The memory element can either be estimated statistically or be measured empirically by dividing the mean log of growth rate over maximum growth rate by the mean log of uptake rate over maximum uptake rate (Eq. 13). Simulations under nitrogen starvation at non-equilibrium in batch culture and statistical comparisons show that our model estimates the growth rate, cell density and resource concentration as well as the complex two-step Droop model. We anticipate that the Monod-memory model may be a good compromise between the limitations of the Monod model and the added complexity and computational challenges of the Droop model.

## Data availability statement

The original contributions presented in the study are included in the article/Supplementary Material. Further details regarding the experimental data, validation results, and the computer codes can be found in public repository on Zenodo. DOI: 10.5281/zenodo.7117583.

## Author contributions

MA and AI designed the study, performed modeling work, and analyzed output data. ZF consulted on writing the manuscript. MA wrote the first draft of the manuscript, and all authors contributed substantially to revisions. All authors contributed to the article and approved the submitted version.

## Funding

This work was supported by the Simons Collaboration on Computational Biogeochemical Modeling of Marine Ecosystems (CBIOMES, grant 549935 to AI).

## Acknowledgments

MA expresses his appreciation for many useful and enlightening discussions with Dr. Joseph Siddons, post-doctoral researcher in Marine Microbial Macroecology Lab at Dalhousie University, and Moein Khalighi of the University of Turku for conversations on fractional calculus.

## Conflict of interest

The authors declare that the research was conducted in the absence of any commercial or financial relationships that could be construed as a potential conflict of interest.

## Publisher's note

All claims expressed in this article are solely those of the authors and do not necessarily represent those of their affiliated organizations, or those of the publisher, the editors and the reviewers. Any product that may be evaluated in this article, or claim that may be made by its manufacturer, is not guaranteed or endorsed by the publisher.

## Supplementary material

The Supplementary Material for this article can be found online at: https://www.frontiersin.org/articles/10.3389/fmars.2022.963734/full#supplementary-material

# References

Akoglu, E. (2020). Simulating nutrient uptake dynamics in plankton models: A case study for the cilician basin marine ecosystem. *Turkish J. Fish. Aquat. Sci.* 20, 603–612. doi: 10.4194/1303-2712-v20_8_03

Amirian, M. M., Towers, I., Jovanoski, Z., and Irwin, A. J. (2020). Memory and mutualism in species sustainability: A time-fractional lotka-volterra model with harvesting. *Heliyon* 6, e04816. doi: 10.1016/j.heliyon.2020.e04816

Anugerahanti, P., Kerimoglu, O., and Smith, S. L. (2021). Enhancing ocean biogeochemical models with phytoplankton variable composition. *Front. Mar. Sci.* 8, 944. doi: 10.3389/fmars.2021.675428

Aumont, O., Eth´ e´, C., Tagliabue, A., Bopp, L., and Gehlen, M. (2015). PISCES-v2: an ocean biogeochemical model for carbon and ecosystem studies. *Geosci. Model. Dev.* 8, 2465–2513. doi: 10.5194/gmd-8-2465-2015

Boethling, R. S., and Alexander, M. (1979). Effect of concentration of organic chemicals on their biodegradation by natural microbial communities. *Appl. Environ. Microbiol.* 37, 1211–1216. doi: 10.1128/aem.37.6.1211-1216.1979

Burmaster, D. E. (1979). The continuous culture of phytoplankton: mathematical equivalence among three steady-state models. *Am. Nat.* 113, 123–134. doi: 10.1086/283368

Droop, M. R. (1968). Vitamin B12 and marine ecology. IV. the kinetics of uptake, growth and inhibition in monochrysis lutheri. *J. Mar. Biol. Assoc. United Kingdom* 48, 689–733. doi: 10.1017/S0025315400019238

Dutkiewicz, S., Cermeno, P., Jahn, O., Follows, M. J., Hickman, A. E., Taniguchi, D. A., et al. (2020). Dimensions of marine phytoplankton diversity. *Biogeosciences* 17, 609–634. doi: 10.5194/bg-17-609-2020

Eftekhari, L., and Amirian, M. M. (2022). Stability analysis of fractional order memristor synapse-coupled hopfield neural network with ring structure. *Cogn. Neurodynamics*, 1–15. doi: 10.1007/s11571-022-09844-9

Fasham, M. J., Ducklow, H. W., and McKelvie, S. M. (1990). A nitrogen-based model of plankton dynamics in the oceanic mixed layer. *J. Mar. Res.* 48, 591–639. doi: 10.1357/002224090784984678

Finkel, Z. V., Beardall, J., Flynn, K. J., Quigg, A., Rees, T. A. V., and Raven, J. A. (2010). Phytoplankton in a changing world: cell size and elemental stoichiometry. *J. Plankton Res.* 32, 119–137. doi: 10.1093/plankt/fbp098

Finkel, Z., Follows, M., and Irwin, A. (2016). Size-scaling of macromolecules and chemical energy content in the eukaryotic microalgae. *J. Plankton Res.* 38, 1151–1162. doi: 10.1093/plankt/fbw057

Finkel, Z. V., Sebbo, J., Feist-Burkhardt, S., Irwin, A., Katz, M., Schofield, O., et al. (2007). A universal driver of macroevolutionary change in the size of marine phytoplankton over the cenozoic. *Proc. Natl. Acad. Sci.* 104, 20416–20420. doi: 10.1073/pnas.0709381104

Flynn, K. J. (2010). Ecological modelling in a sea of variable stoichiometry: dysfunctionality and the legacy of redfield and monod. *Prog. Oceanogr.* 84, 52–65. doi: 10.1016/j.pocean.2009.09.006

Follett, C. L., Dutkiewicz, S., Ribalet, F., Zakem, E., Caron, D., Armbrust, E. V., et al. (2022). Trophic interactions with heterotrophic bacteria limit the range of prochlorococcus. *Proc. Natl. Acad. Sci.* 119, e2110993118. doi: 10.1073/pnas.2110993118

Follows, M. J., and Dutkiewicz, S. (2011). Modeling diverse communities of marine microbes. *Annu. Rev. Mar. Sci.* 3, 427–451. doi: 10.1146/annurev-marine-120709-142848

Follows, M. J., Dutkiewicz, S., Grant, S., and Chisholm, S. W. (2007). Emergent biogeography of microbial communities in a model ocean. *Science* 315, 1843–1846. doi: 10.1126/science.1138544

Geider, R. J., MacIntyre, H. L., and Kana, T. M. (1998). A dynamic regulatory model of phytoplanktonic acclimation to light, nutrients, and temperature. *Limnol. Oceanogr.* 43, 679–694. doi: 10.4319/lo.1998.43.4.0679

Grover, J. P. (1991). Resource competition in a variable environment: phytoplankton growing according to the variable-internal-stores model. *Am. Nat.* 138, 811–835. doi: 10.1086/285254

Hellweger, F. L., and Kianirad, E. (2007). Accounting for intrapopulation variability in biogeochemical models using agent-based methods. *Environ. Sci. Technol.* 41, 2855–2860. doi: 10.1021/es062046j

Henriques, J., and Balsa-Canto, E. (2021). The monod model is insufficient to explain biomass growth in nitrogen-limited yeast fermentation. *Appl. Environ. Microbiol.* 87, e01084–e01021. doi: 10.1128/AEM.01084-21

Henson, S. A., Cael, B., Allen, S. R., and Dutkiewicz, S. (2021). Future phytoplankton diversity in a changing climate. *Nat. Commun.* 12, 1–8. doi: 10.1038/s41467-021-25699-w

Irwin, A. J., and Finkel, Z. V. (2018). "Phytoplankton functional types: a trait perspective. chap. 11," in *Microbial ecology of the oceans*. Eds. J. M. Gasol and D. L. Kirchman (Springer: Cold Spring Harbor Laboratory) *bioRxiv*. 435.

Irwin, A. J., Finkel, Z. V., Muller-Karger, F. E., and Ghinaglia, L. T. (2015). Phytoplankton adapt to changing ocean environments. *Proc. Natl. Acad. Sci.* 112, 5762–5766. doi: 10.1073/pnas.1414752112

Irwin, A. J., Finkel, Z. V., Schofield, O. M., and Falkowski, P. G. (2006). Scaling-up from nutrient physiology to the size-structure of phytoplankton communities. *J. Plankton Res.* 28, 459–471. doi: 10.1093/plankt/fbi148

Khalighi, M., Sommeria-Klein, G., Gonze, D., Faust, K., and Lahti, L. (2022). Quantifying the impact of ecological memory on the dynamics of interacting communities. *PloS Comput. Biol.* 18, e1009396. doi: 10.1371/journal.pcbi.1009396

Kovarova-Kovar, K., and Egli, T. (1998). Growth kinetics of suspended microbial cells: from singlesubstrate-controlled growth to mixed-substrate kinetics. *Microbiol. Mol. Biol. Rev.* 62, 646–666. doi: 10.1128/MMBR.62.3.646-666.1998

Kwiatkowski, L., Aumont, O., Bopp, L., and Ciais, P. (2018). The impact of variable phytoplankton stoichiometry on projections of primary production, food quality, and carbon uptake in the global ocean. *Global Biogeochem. Cycles* 32, 516–528. doi: 10.1002/2017GB005799

Lee, E., Jalalizadeh, M., and Zhang, Q. (2015). Growth kinetic models for microalgae cultivation: a review. *Algal Res.* 12, 497–512. doi: 10.1016/j.algal.2015.10.004

Liefer, J. D., Garg, A., Fyfe, M. H., Irwin, A. J., Benner, I., Brown, C. M., et al. (2019). The macromolecular basis of phytoplankton c: N: P under nitrogen starvation. *Front. Microbiol.* 10, 763. doi: 10.3389/fmicb.2019.00763

Marr, A. G., Nilson, E., and Clark, D. (1963). The maintenance requirement of escherichia coli. *Ann. New York Acad. Sci.* 102, 536–548. doi: 10.1111/j.1749-6632.1963.tb13659.x

Matlob, M. A., and Jamali, Y. (2019). The concepts and applications of fractional order differential calculus in modeling of viscoelastic systems: A primer. *Crit. Reviews™ Biomed. Eng.* 47, 249–276. doi: 10.1615/CritRevBiomedEng.2018028368

Mei, Z. P., Finkel, Z. V., and Irwin, A. J. (2009). Light and nutrient availability affect the size-scaling of growth in phytoplankton. *J. Theor. Biol.* 259, 582–588. doi: 10.1016/j.jtbi.2009.04.018

Moore, J. K., Doney, S. C., Kleypas, J. A., Glover, D. M., and Fung, I. Y. (2001). An intermediate complexity marine ecosystem model for the global domain. *Deep Sea Res. Part II: Topical Stud. Oceanogr.* 49, 403–462. doi: 10.1016/S0967-0645(01)00108-4

Morel, F. M. (1987). Kinetics of nutrient uptake and growth in phytoplankton 1. *J. Phycol.* 23, 137–150. doi: 10.1111/j.1529-8817.1987.tb04436.x

Nicholson, D., Stanley, R., and Doney, S. (2018). A phytoplankton model for the allocation of gross photosynthetic energy including the trade-offs of diazotrophy. *J. Geophys. Res.: Biogeosci.* 123, 1796–1816. doi: 10.1029/2017JG004263

Nocedal, J., and Wright, S. (2006). *Numerical optimization* (Springer Science & Business Media).

Pahlow, M., and Oschlies, A. (2013). Optimal allocation backs droop's cell-quota model. *Mar. Ecol. Prog. Ser.* 473, 1–5. doi: 10.3354/meps10181

Pirt, S. (1965). The maintenance energy of bacteria in growing cultures. *Proc. R. Soc. London Ser. B. Biol. Sci.* 163, 224–231. doi: 10.1098/rspb.1965.0069

Podlubny, I. (2002). Geometric and physical interpretation of fractional integration and fractional differentiation. *Fractional Calculus Appl. Anal.* 5, 367–386. doi: 10.48550/arXiv.math/0110241

Rittmann, B. E., and McCarty, P. L. (1980). Evaluation of steady-state-biofilm kinetics. *Biotechnol. Bioeng.* 22, 2359–2373. doi: 10.1002/bit.260221111

Sancho, M. M., Castillo, J. J., and El Yousfi, F. (1997). Influence of phosphorus concentration on the growth kinetics and stoichiometry of the microalga scenedesmus obliquus. *Process Biochem.* 32, 657–664. doi: 10.1016/S0032-9592 (97)00017-4

Shuter, B. (1978). Size dependence of phosphorus and nitrogen subsistence quotas in unicellular microorganisms 1. *Limnol. Oceanogr.* 23, 1248–1255. doi: 10.4319/lo.1978.23.6.1248

Shuter, B. (1979). A model of physiological adaptation in unicellular algae. *J. Theor. Biol.* 78, 519–552. doi: 10.1016/0022-5193(79)90189-9

Smith, S., Pahlow, M., Merico, A., Acevedo-Trejos, E., Sasai, Y., Yoshikawa, C., et al. (2016). Flexible phytoplankton functional type (flexpft) model: size-scaling of traits and optimal growth. *J. Plankton Res.* 38, 977–992. doi: 10.1093/plankt/fbv038

Smith, S., Yamanaka, Y., Pahlow, M., and Oschlies, A. (2009). Optimal uptake kinetics: physiological acclimation explains the pattern of nitrate uptake by phytoplankton in the ocean. *Mar. Ecol. Prog. Ser.* 384, 1–12. doi: 10.3354/meps08022

Soetaert, K., Petzoldt, T., and Setzer, R. W. (2010). Solving differential equations in r: Package deSolve. *J. Stat. Softw.* 33, 1–25. doi: 10.18637/jss.v033.i09

Tozzi, S., Schofield, O., and Falkowski, P. (2004). Historical climate change and ocean turbulence as selective agents for two key phytoplankton functional groups. *Mar. Ecol. Prog. Ser.* 274, 123–132. doi: 10.3354/meps274123

Tsakalakis, I., Follows, M. J., Dutkiewicz, S., Follett, C. L., and Vallino, J. J. (2022). Diel light cycles affect phytoplankton competition in the global ocean. *Glob. Ecol. Biogeogr* 31 (9), 1838–1849. doi: 10.1101/2021.05.19.444874

Van Uden, N. (1967). Transport-limited growth in the chemostat and its competitive inhibition; a theoretical treatment. *Archiv. fur¨ Mikrobiologie* 58, 145–154. doi: 10.1007/BF00406675

Verdy, A., Follows, M., and Flierl, G. (2009). Optimal phytoplankton cell size in an allometric model. *Mar. Ecol. Prog. Ser.* 379, 1–12. doi: 10.3354/meps07909

Ward, B. A., Marañón, E., Sauterey, B., Rault, J., and Claessen, D. (2017). The size dependence of phytoplankton growth rates: A trade-off between nutrient uptake and metabolism. *Am. Nat.* 189, 170–177. doi: 10.1086/689992

Wu, Y., Campbell, D. A., Irwin, A. J., Suggett, D. J., and Finkel, Z. V. (2014). Ocean acidification enhances the growth rate of larger diatoms. *Limnol. Oceanogr.* 59, 1027–1034. doi: 10.4319/lo.2014.59.3.1027